\documentclass[final,5p,times,twocolumn]{elsarticle}
\usepackage{amssymb}
\usepackage{graphicx}
\usepackage{caption}
\usepackage{subcaption}
\usepackage{amsmath}
\usepackage{amssymb}
\usepackage{tikz}
\usepackage{secdot}
\usepackage{epstopdf}
\usepackage{geometry}
\usepackage{hyperref}
\usepackage{tabularx,ragged2e,booktabs,caption}
\usepackage{cite}
\usepackage{float}
\usepackage{lipsum}
\usepackage{fixltx2e}
\usepackage{multirow}
\biboptions{numbers,sort&compress}
\journal{Intermetallics}

\begin{document}

\begin{frontmatter}

\title{Electric field gradients in Zr$_8$Ni$_{21}$ and Hf$_8$Ni$_{21}$ intermetallic compounds;
Results from perturbed angular correlation measurements and
first-principles density functional theory 
}
\author{S.K. Dey$^1$}
\ead{skumar.dey@saha.ac.in}
\author{C.C. Dey$^1$\corref{cor1}}
\cortext[cor1]{corresponding author}
\ead{chandicharan.dey@saha.ac.in}
\author{S. Saha$^1$}
\ead{satyajit.saha@saha.ac.in}
\author{J. Belo$\check{\text{s}}$evi\'{c}-$\check{\text{C}}$avor$^2$}
\ead{cjeca@vin.bg.ac.rs}
\address{$^1$Saha Institute of Nuclear Physics; 1/AF, Bidhannagar, Kolkata - 700 064, India}
\address{$^2$Institute of Nuclear Sciences Vinca, University of Belgrade, P. O. Box 522,\\ 11001 Belgrade, Serbia}

\begin{abstract}
Numerous technological applications of Ni-based Zr and Hf
intermetallic alloys promoted comprehensive studies in Zr$_8$Ni$_{21}$ and Hf$_8$Ni$_{21}$ by perturbed angular correlation (PAC) spectroscopy, which were not
studied earlier until this report. The different phases produced in the samples
have been identified by PAC and X-ray diffraction (XRD) measurements. Using $^{181}$Hf
probe, two non-equivalent Zr/Hf sites have been observed in both Zr$_8$Ni$_{21}$ and Hf$_8$Ni$_{21}$
compounds. From present PAC measurements in Zr$_8$Ni$_{21}$,
a component due to the production of Zr$_7$Ni$_{10}$ by eutectic reaction from the
liquid metals is also observed. The phase Zr$_7$Ni$_{10}$, however, is not found from the
XRD measurement. In Zr$_8$Ni$_{21}$, while the results do not change appreciably
up to 973 K exhibit drastic changes at 1073 K. In Hf$_8$Ni$_{21}$, similar results for the two non-equivalent sites have
been found but site fractions are in reverse order. In this alloy, a different
contaminating phase, possibly due to HfNi$_3$, has been found from PAC measurements but is not found
from XRD measurement. Density functional
theory (DFT) based calculations of electric field gradient (EFG) and asymmetry parameter
($\eta$) at the sites of $^{181}$Ta probe nucleus allowed us to assign the observed EFG fractions to the various
lattice sites in (Zr/Hf)$_8$Ni$_{21}$ compounds. 
\end{abstract}


\end{frontmatter}

\section{Introduction}
 Intermetallic binary alloys between the two
 elements of A (Hf/Zr/Ti) and B (Fe/Co/Ni)
 have many technological applications. They
 have properties like superior strength, corrosion
 resistance, hydrogen storage capacity, ferromagnetism
 and have applications in space craft industry,
 fuel cell etc. The Zr-Ni alloys, particularly,
 have excellent hydrogen storage properties and
 have applications in Ni-metal hydride (MH)
 rechargeable batteries as a negative electrode \citep{JMJoubert,Ruiz,FCRuiz}. The gaseous hydrogen storage characteristics
 of four intermetallic compounds in the Zr-Ni
 system viz. Zr$_9$Ni$_{11}$, Zr$_7$Ni$_{10}$, Zr$_8$Ni$_{21}$ and
 Zr$_2$Ni$_7$ were compared by Joubert et al. \citep{JMJoubert}.
 It was found that hydrogen storage
 capacities in hydrogen atoms per metal atom
 (H/M) for these four compounds are
 0.93 (10 bar), 1.01 (10 bar), 0.34 (25 bar) and 0.29 (25 bar), respectively,
 at room temperature and the storage capacities for Zr$_8$Ni$_{21}$,
 Zr$_2$Ni$_7$ are completely reversible.
 The hydrogen reversible capacities for
 Zr$_9$Ni$_{11}$ and Zr$_7$Ni$_{10}$, are 50\% and 77\%, respectively.
 The electrochemical properties of Zr$_7$Ni$_{10}$,
 Zr$_9$Ni$_{11}$ and Zr$_8$Ni$_{21}$ were studied by Ruiz et al.
 \citep{Ruiz,FCRuiz} and Nei et al. \citep{Regmi,S.O.Salley}. It was found \citep{FCRuiz}
 that Zr$_8$Ni$_{21}$ alloy had a better charge/discharge
 performance than Zr$_7$Ni$_{10}$ and Zr$_9$Ni$_{11}$. A detailed
 review on the electrochemical properties of various
 compounds of Zr-Ni system for application of these materials
 in Ni-MH battery was reported by Young et al. \citep{Young}.
 The intermetallic compound ZrNi$_5$ was reported to
 have strong ferromagnetic properties by
 Drulis et al. \citep{Drulis}. A. Amamou \citep{Amamou} reported the
 electronic structure of various compounds in
 the Zr-Ni system, namely, Zr$_2$Ni, ZrNi, Zr$_8$Ni$_{21}$
 and ZrNi$_5$. Both magnetic and
 structural properties of the A$_x$B$_y$ compounds
 can be studied experimentally by the nuclear technique of perturbed
 angular correlation (PAC), which uses a suitable
 radioactive isotope (usually $^{181}$Hf) to characterize the materials. Using this technique, 
 different intermetallic
 compounds in the Zr-Ni system have recently 
 been studied, namely, ZrNi$_5$ \citep{Dey,SilvaJMMM}, Zr$_2$Ni$_7$ \citep{CCDeyPhysica} and ZrNi \citep{CCDeyJPCS}.
 From the studies in ZrNi$_5$ system \citep{Dey,SilvaJMMM}, however, no magnetic
 interaction was found, in contradiction with the result from previous measurements by Drulis et al. \citep{Drulis}. 
 
 Considering the important applications
 of Zr$_8$Ni$_{21}$, Zr$_9$Ni$_{11}$ and Zr$_7$Ni$_{10}$ as described above,
 we have done detailed PAC measurements in Zr$_8$Ni$_{21}$
 and Hf$_8$Ni$_{21}$ intermetallic compounds. To the best of our knowledge,
 there is no measurement by PAC technique to characterize
 these materials. From previous studies \citep{Shen}, a pure
 single phase Zr$_8$Ni$_{21}$ was not found to be produced by
 arc melting preparation. It was found \citep{Shen} that
 Zr$_8$Ni$_{21}$ was not formed congruently from the
 liquid. The Zr$_2$Ni$_7$ was first solidified from the liquid
 and then reacted with the remaining liquid to form
 Zr$_8$Ni$_{21}$ alloy peritectically. The Zr$_7$Ni$_{10}$ was formed eutectically from Zr$_8$Ni$_{21}$. The two other phases
 viz. Zr$_2$Ni$_7$ and Zr$_7$Ni$_{10}$ were produced along with
 Zr$_8$Ni$_{21}$ and were confirmed by scanning electron
 microscopy (SEM)/X-ray energy dispersive spectroscopy (EDS)
 compositional mapping and transmission electron
 microscopy (TEM) \citep{Shen}. However,
 in the present report, we have studied both Zr$_8$Ni$_{21}$ and
 Hf$_8$Ni$_{21}$ to
 identify and characterize the different phases produced in these compounds. The secondary
 phases of small fractions that are produced along with the main
 phase can be determined quite accurately by this technique. The structural and compositional stability of Zr$_8$Ni$_{21}$/Hf$_8$Ni$_{21}$ phases have been 
 studied by temperature dependent PAC measurements.
 
 The crystal structure of Zr$_8$Ni$_{21}$ is known to be
 triclinic \citep{JM} and is isotropic to that of Hf$_8$Ni$_{21}$ \citep{Bsenko}.
 The lattice parameters of Zr$_8$Ni$_{21}$ are
 $a$=6.476 \AA, $b$=8.064 \AA,
 $c$=8.594 \AA, $\alpha$=75.15$^\circ$, $\beta$=68.07$^\circ$ and $\gamma$=75.23$^\circ$ as
 determined by X-ray diffraction analysis.
 
 In the PAC technique \citep{Schatz,Catchen,Zacate}, the angular correlation
 of a $\gamma$-$\gamma$ cascade in a suitable probe nucleus is perturbed by the interaction of the probe nuclear moments 
 with the electric
 field gradients/magnetic fields generated at
 the probe site due to surrounding charge distribution.
 The crystalline electric field gradient (EFG)
 and the internal magnetic field in a
 magnetic material can be determined by the PAC
 technique if the values of electromagnetic moments
 of the intermediate level of the probe nucleus are known. As the EFG depends on the charge distribution of the probe-nucleus 
 environment, the temperature evolution of the 
 lattice properties such as crystallographic structure, 
 imperfections or defects, can be monitored by applying PAC technique over a wide temperature range. 
 The combination of PAC measurements and ab-initio calculations proved to be an excellent method
 to study the structural phase stabilities and the localization of the impurities in the host lattice \citep{Errico, Wodniecki}.
In this paper, results of temperature dependent PAC measurements (77-1073 K) in both Zr$_8$Ni$_{21}$ and Hf$_8$Ni$_{21}$ as well
as DFT calculations are reported. The calculated EFG values at the $^{181}$Ta impurity sites are compared with experimental results.

\section{Experimental details}

The intermetallic compounds Zr$_8$Ni$_{21}$ and Hf$_8$Ni$_{21}$ were prepared by arc melting in an argon atmosphere. Stoichiometric amounts of high purity metals 
procured from M/S Alfa Aesar were used to prepare the samples. The purity of Zr (excluding Hf), Hf (excluding Zr) and Ni metals used were 99.2\%, 
99.95\% and 99.98\% respectively. For each sample, the constituent metals were alloyed homogeneously by repeated melting and then
activated 
by remelting with a piece of $^{181}$Hf metal. The $^{181}$Hf metal was prepared by thermal neutron capture of natural $^{180}$Hf metal 
in the Dhruba reactor, Mumbai using a flux $\sim$10$^{13}$/cm$^2$/s. In both cases, shiny globule samples were formed and these 
were then sealed in evacuated quartz tubes for high temperature measurements. Different inactive samples of Zr$_8$Ni$_{21}$ and Hf$_8$Ni$_{21}$ were
prepared similarly 
for XRD measurements. The X-ray powder diffraction measurements have been carried out using the Rigaku X-ray diffractometer TTRAX-III and Cu K$_\alpha$ radiation.

The TDPAC technique measures the effect of perturbations of $\gamma$-$\gamma$ angular correlation of the probe nucleus through the hyperfine 
interaction. In the present case, the probe $^{181}$Hf substitutes the Zr atom in Zr$_8$Ni$_{21}$ and is a constituent element in Hf$_8$Ni$_{21}$. 
In the $\beta^-$ decay of $^{181}$Hf, it
populates the 615 keV excited level of $^{181}$Ta which emits two successive $\gamma$ rays, 133 and 482 keV
passing through the 482 keV level with a half-life 10.8 ns and a spin angular momentum $I$=5/2$^+\hbar$ \citep{Firestone}. The angular 
correlation of the 133-482 keV cascade is perturbed by the extranuclear electric field gradients.
 
The perturbation function $G_2(t)$ for the electric quadrupole interaction in a polycrystalline material is given by \citep{Schatz} 
 \begin{equation}
  G_2(t)=S_{20}(\eta) + \sum^{3}_{i=1}S_{2i}(\eta)cos(\omega_it)exp(-\delta\omega_it)exp\big[\frac{-(\omega_i\tau_R)^2}{2}\big]
 \label{eqn:Stokes}
 \end{equation}
The frequencies $\omega_i$ correspond to transitions between the sub-levels of the intermediate state that arise due to 
nuclear quadrupole interaction (NQI). 
The parameter $\delta$ is the
frequency distribution width (Lorentzian damping) which takes care of the chemical inhomogeneities 
in the sample and $\tau_R$ is the time resolution of the coincidence set up. Due to the presence of 
various non-equivalent sites, the perturbation factor $G_2$(t) can generally be expressed as 
  \begin{equation}
 G_2(t)=\sum_if_iG^{i}_{2}(t),
  \label{eqn:Bhatnagar}
 \end{equation}
  where $f_i$ is the site fraction of the
  $i$-th component. 
A fitting to eqn. ($\ref{eqn:Stokes}$) determines the maximum component $V_{zz}$ of the 
electric field gradient from the measured quadrupole frequency $\omega_Q$ given by
\begin{equation}
\omega_Q= \frac{eQV_{zz}}{4I(2I-1)\hbar},
 \label{eqn:raman}
\end{equation}
where $Q$ is the nuclear quadrupole moment of the 482 keV intermediate state (2.36 b). 
For an axially symmetric EFG ($\eta=0$), $\omega_Q$ is related to $\omega_1$, $\omega_2$ and $\omega_3$ by 
\begin{equation}
 \omega_Q=\omega_1/6=\omega_2/12=\omega_3/18.
 \label{eqn:prafulla}
\end{equation}
The principal EFG components obey the relations
\begin{equation}
V_{xx} + V_{yy} + V_{zz}=0 \quad \text{and}\quad
V_{zz}\ge V_{yy}\ge V_{xx}.
 \label{eqn:hizenberg}
\end{equation} 
 The EFG can therefore be designated by two parameters only viz. $V_{zz}$ and $\eta$. The asymmetry parameter $\eta$ is defined as  
     \begin{equation}
  \eta=\frac{(V_{xx}-V_{yy})}{V_{zz}},\quad \text{}\quad 0\le\eta\le1.
  \label{eqn:newton}
 \end{equation} 
 
    The TDPAC spectrometer used for present measurements was a four detector
    LaBr$_3$(Ce)-BaF$_2$ set up with crystal sizes 38$\times$25.4 mm$^2$ for LaBr$_3$(Ce) and 50.8$\times$50.8 mm$^2$ for 
   BaF$_2$. The 133 keV $\gamma$-rays were selected in LaBr$_3$(Ce) detectors. 
   Standard slow-fast
   coincidence assemblies were employed to collect data at 180$^\circ$ and 90$^\circ$. 
   Typical prompt time resolution (FWHM)$\sim$800 ps was obtained for the $^{181}$Hf energy window settings. 
   The perturbation function $G_2$($t$) is found from the four coincidence spectra at 180$^\circ$ and 90$^\circ$.
   Details of the experimental set up and data analysis can be found in reference \citep{pramana}.
  
    \begin{table}[t!]
\begin{center} 
\captionof{table}{\small{ Results of PAC measurements in Zr$_8$Ni$_{21}$}}
\scalebox{0.65}{
\begin{tabular}{ l l l l l l l l } 
 \hline  \\ [-0.9ex]
Temperature (K)  &Component    & $\omega_Q$ (Mrad/s)     & $\eta$     & $\delta$($\%$)   & $f$($\%$)      & Assignment     \\ [1.5ex]
 \hline  \\ 
 
77             &1        & 77.9(4)                 & 0.80(1)          & 2(1)            & 55(3)      & Zr$_8$Ni$_{21}^{(1)}$        \\   
                 &2      & 55.8(7)                 & 0.68(2)          & 0                & 26(3)         & Zr$_7$Ni$_{10}$   \\   
                 &3      & 101.5(9)                 & 0.73(3)          & 0                & 18(3)         & Zr$_8$Ni$_{21}^{(2)}$   \\   \\ 

298             &1        & 75.8(2)                 & 0.77(1)          & 1.1(6)            & 57(3)      & Zr$_8$Ni$_{21}^{(1)}$        \\   
                 &2      & 54.4(4)                 & 0.69(1)          & 0                & 27(3)         & Zr$_7$Ni$_{10}$   \\   
                 &3      & 100.6(8)                 & 0.72(3)          & 0                & 15(3)        & Zr$_8$Ni$_{21}^{(2)}$   \\   \\ 

373               &1     & 74.6(6)                 & 0.75(2)          & 2(1)                & 53(3)    &Zr$_8$Ni$_{21}^{(1)}$           \\   
                 &2      & 53(2)                 & 0.71(3)          & 0                & 26(3)  & Zr$_7$Ni$_{10}$         \\    
                 & 3     & 97(2)                 & 0.76(4)          & 0                & 21(3)  & Zr$_8$Ni$_{21}^{(2)}$        \\ \\
                       
473              &1      & 73.1(5)                 & 0.76(2)          & 2.4(8)                & 63(3)    & Zr$_8$Ni$_{21}^{(1)}$             \\   
                 &2      & 52.1(7)                 & 0.72(4)          & 0                & 37(3)    & Zr$_7$Ni$_{10}$        \\     \\ 
                      
573              &1      & 71.8(3)                 & 0.77(1)          & 2.5(6)                & 62(3)   & Zr$_8$Ni$_{21}^{(1)}$            \\   
                 &2      & 52.9(5)                 & 0.70(2)          & 0                & 38(3)          & Zr$_7$Ni$_{10}$ \\    \\ 
                       
673              &1      & 70.1(6)              & 0.77(2)          & 4.5(8)               & 71(3)  & Zr$_8$Ni$_{21}^{(1)}$              \\  
                 &2      & 50.8(4)                 & 0.72(2)          & 0                & 29(3)   & Zr$_7$Ni$_{10}$       \\      \\   

773                 &1     & 68(1)                 & 0.73(4)        & 4(2)                & 64(3)  & Zr$_8$Ni$_{21}^{(1)}$         \\     
                  &2      & 48.4(9)                & 0.75(5)          & 0                & 36(3)   & Zr$_7$Ni$_{10}$           \\  \\
                       
873              &1      & 67(3)                 & 0.79(10)          & 8(4)                & 72(3)    & Zr$_8$Ni$_{21}^{(1)}$       \\   
                 &2      & 48.8(8)                & 0.73(5)         & 0                & 28(3)  &   Zr$_7$Ni$_{10}$          \\ \\        
                     
973                 &1      & 65(2)                 & 0.80(7)          & 7(3)                & 71(3)           &Zr$_8$Ni$_{21}^{(1)}$\\ 
                   &2      & 47.3(6)                & 0.74(4)          & 0                & 29(3)      &Zr$_7$Ni$_{10}$    \\       \\ 

1073               &1     & 52.8(2)                 & 0          & 1.3(7)               & 68(3) & Hf    \\   
                   &2     & 47(1)                 & 0.71(9)          & 5(3)                & 32(3)            &  Zr$_7$Ni$_{10}$     \\   \\
                       
298$^*$        &1    & 56.8(2)                 & 0         & 4.0(9)                & 70(3)         & Hf  \\ 
                    &2   & 74.4(8)                & 0.79(1)          & 0                & 23(3)    & Zr$_8$Ni$_{21}^{(1)}$    \\   
                    &3   & 74(2)                & 0.21(10)          & 0                & 7(3)    &  Zr$_2$Ni$_7$   \\                                               

\hline
\end{tabular}}
\begin{flushleft}
 \small{$^*$ after measurement at 1073 K}\\
\end{flushleft}
\label{tab:Zr8Ni21table}
\end{center}
\end{table}

  \begin{figure*}[t!]
  \centering
  \begin{subfigure}[t]{0.45\textwidth}
\centering
\includegraphics[scale=.28]{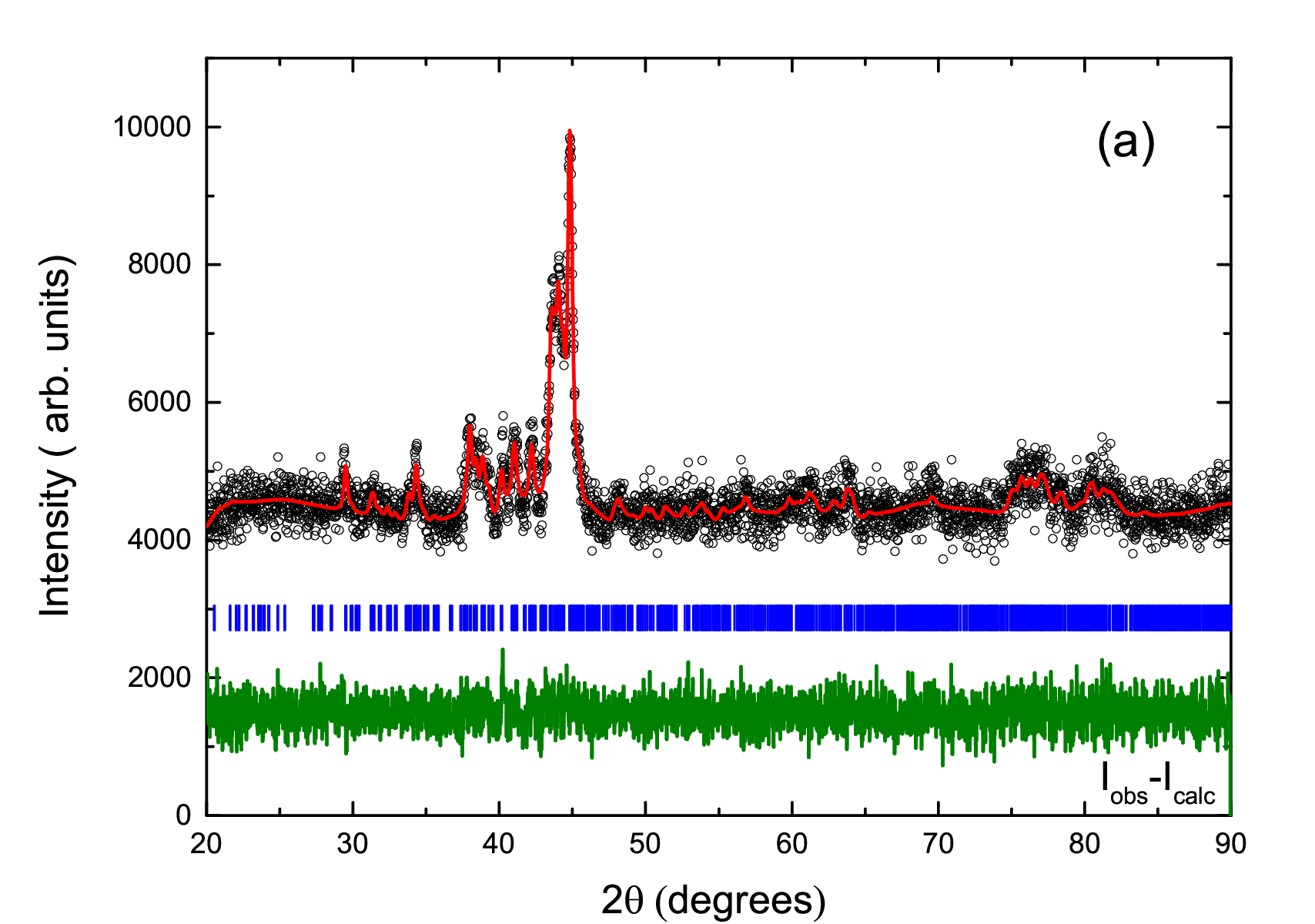}
\label{XRD_Zr8Ni21}
\end{subfigure}
  \begin{subfigure}[t]{0.45\textwidth}
\centering
\includegraphics[scale=.28]{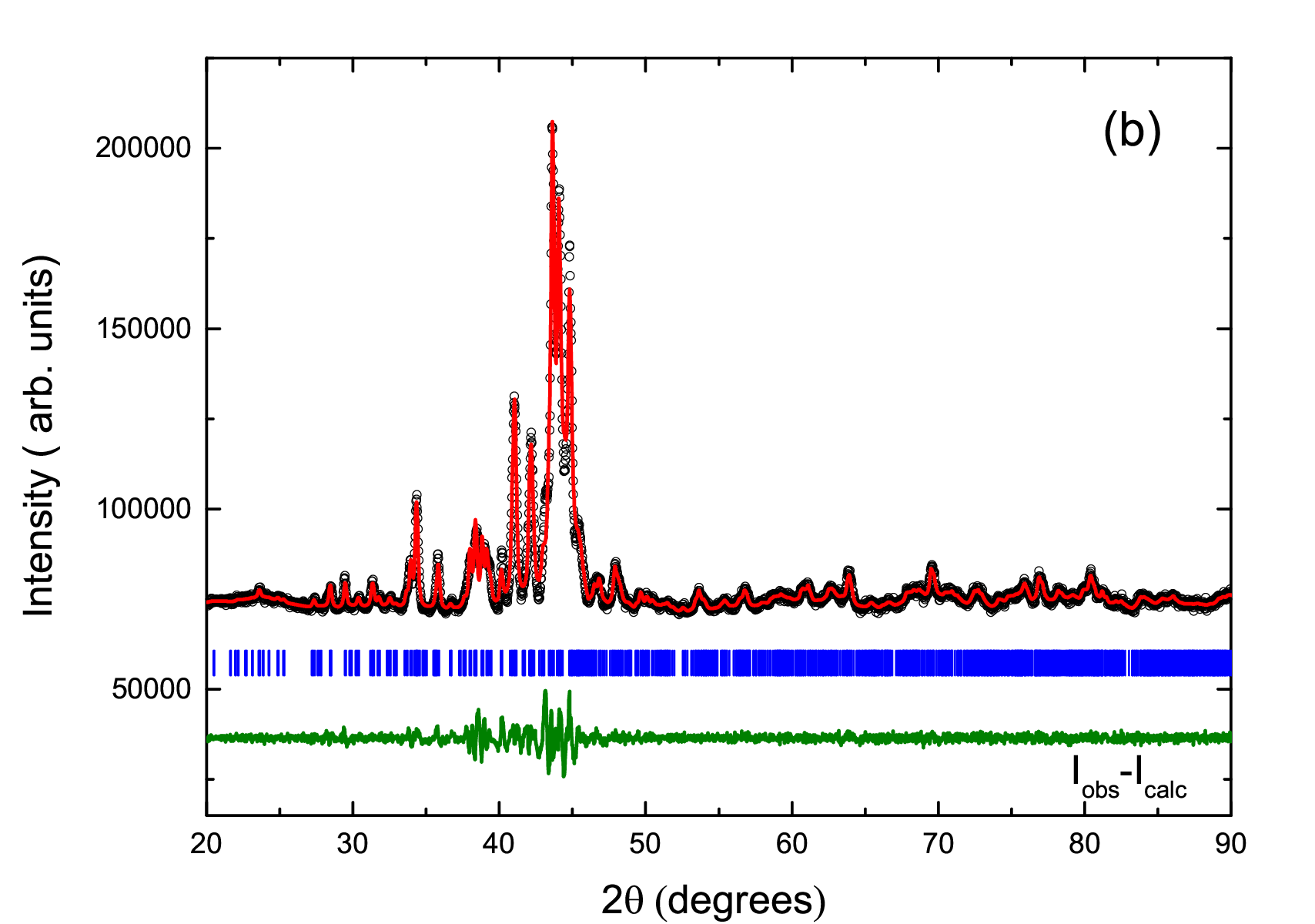}

\end{subfigure}
\caption{ The XRD spectra in Zr$_8$Ni$_{21}$. Figure (a) shows the spectrum in as prepared sample and figure (b) shows the spectrum in a 
sample annealed at 1073 K for two days. The line represents the fit to the measured data, 
the vertical bars denote the Bragg angles and
the bottom line shows the difference between the observed and the fitted pattern. }
\label{XRD_Zr8Ni21}
\end{figure*}

\begin{figure*}
\begin{center}
\includegraphics[scale=.65]{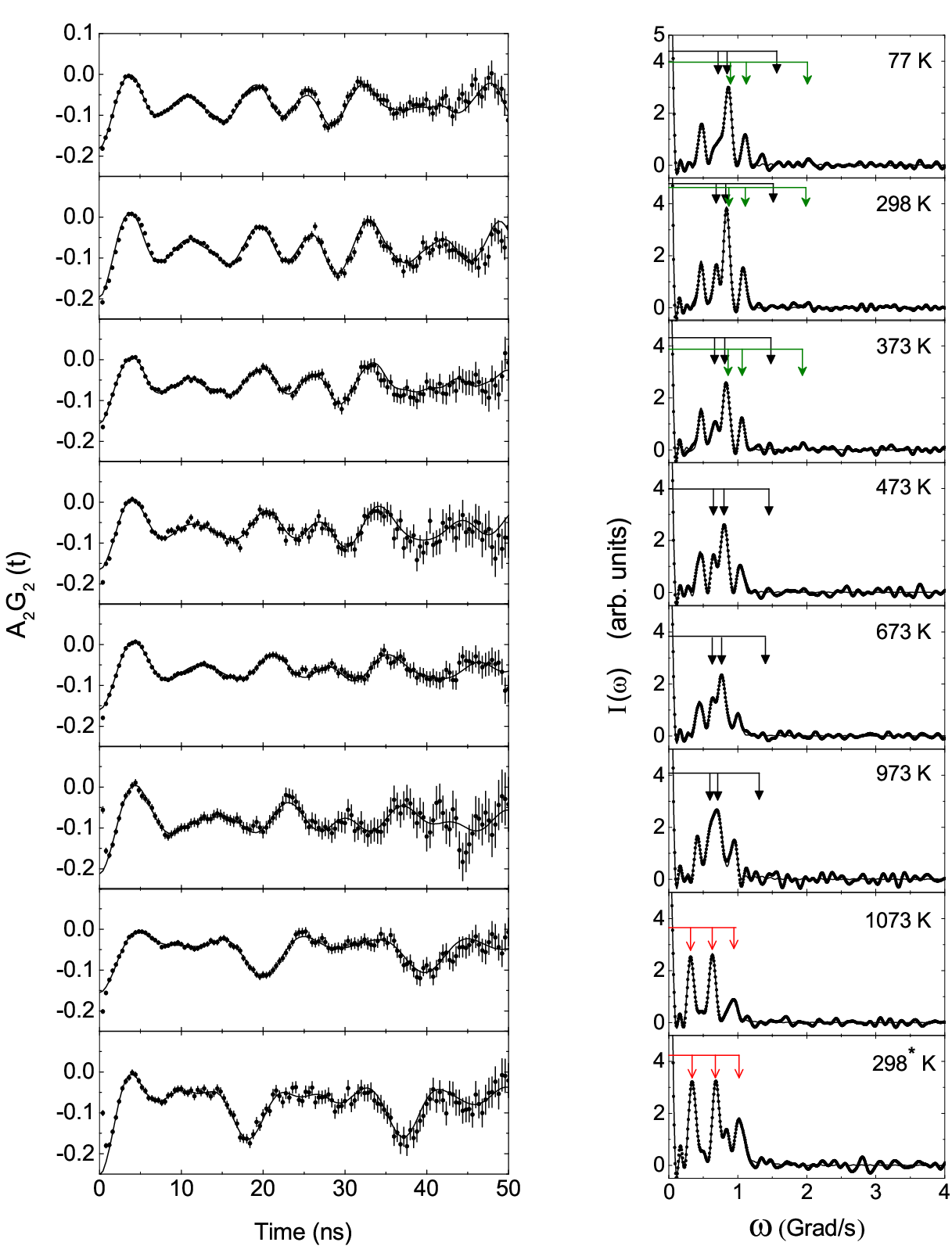}
\end{center}
\caption{\small{TDPAC spectra in Zr$_8$Ni$_{21}$ at different temperatures. Left panel shows the time spectra and
the right panel shows the corresponding Fourier transforms. The PAC spectrum designated
by 298$^*$ K is taken after the measurement at 1073 K. The two sets of arrows in each
Fourier spectrum (up to 373 K) correspond to two non-equivalent $^{181}$Ta sites
in Zr$_8$Ni$_{21}$. Arrows shown in the Fourier spectra at 1073 and 298$^*$ K correspond to Hf.}}
\label{Zr8Ni21TDPAC}
\end{figure*}

\begin{figure*}
\begin{center}
\includegraphics[scale=.25]{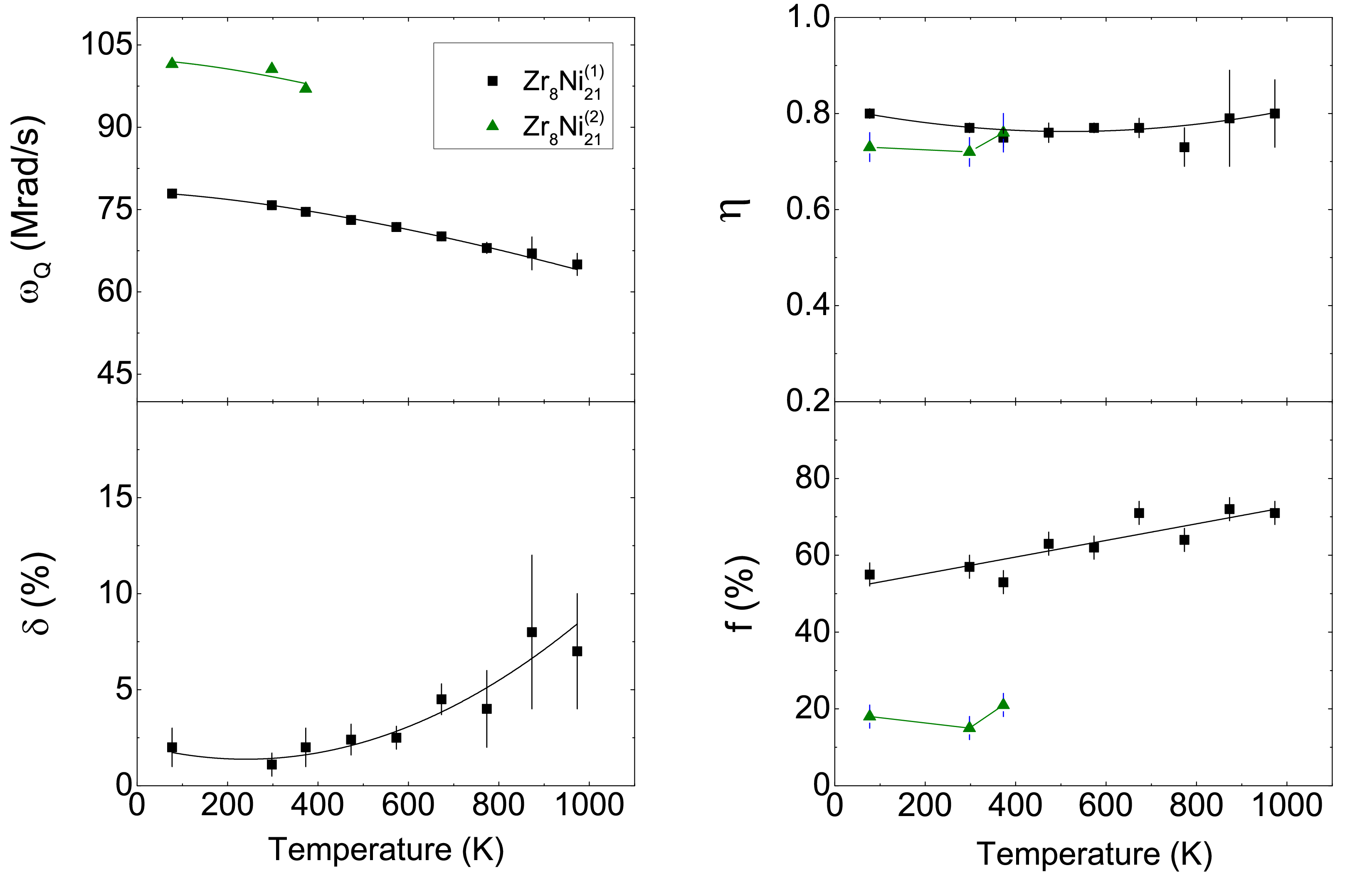}
\end{center}
\caption{Variations of quadrupole frequency ($\omega_Q$), asymmetry parameter ($\eta$) and site fraction $f$(\%)
with temperature for the two non-equivalent $^{181}$Ta sites in Zr$_8$Ni$_{21}$. Variation of $\delta$ is shown for 
the component Zr$_8$Ni$_{21}^{(1)}$.}
\label{Zr8Ni21parameter}
\end{figure*}

 \section{PAC results}
\begin{figure*}
\begin{center}
\includegraphics[scale=.40]{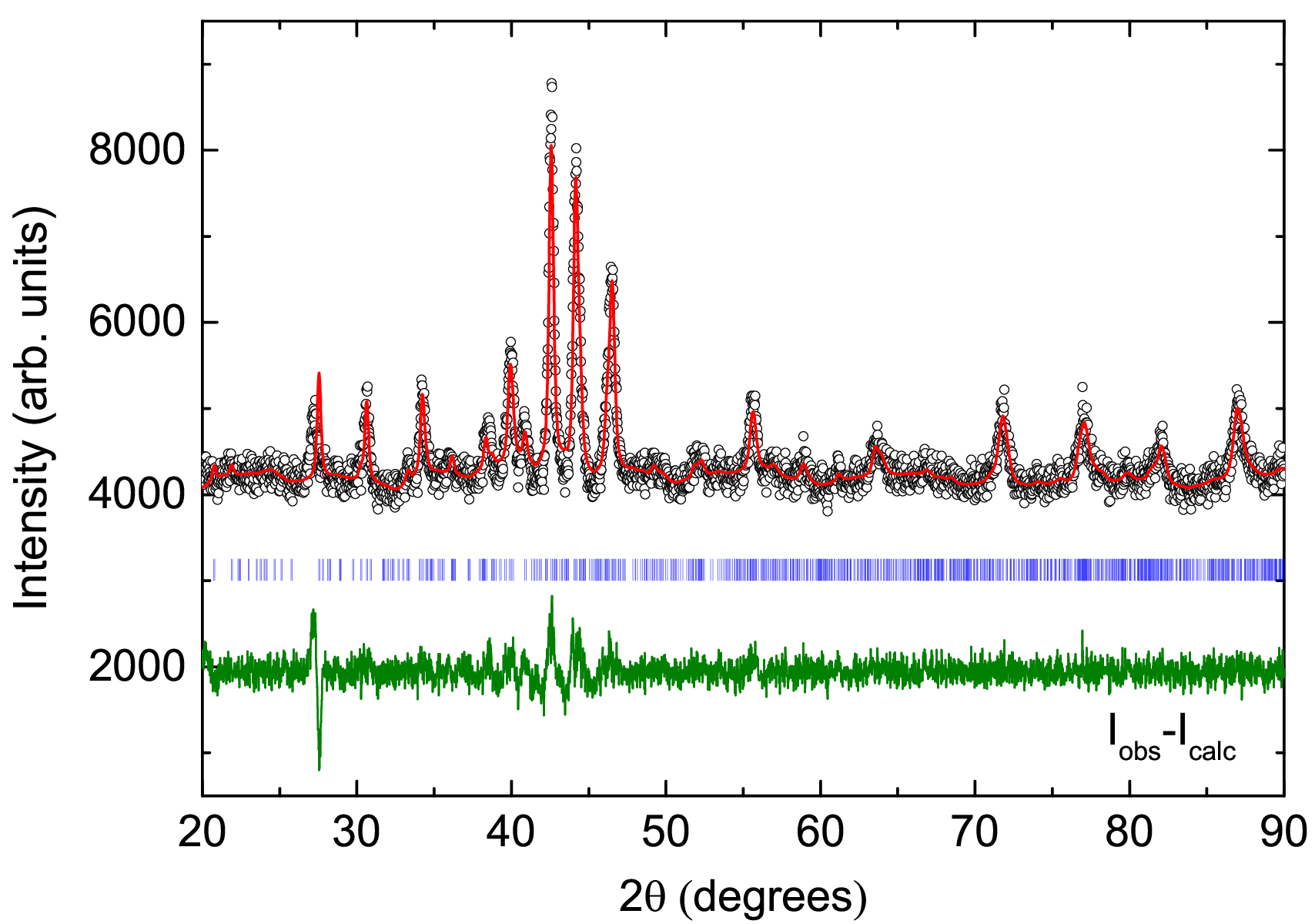}
\end{center}
\caption{The background subtracted XRD powder pattern in Hf$_8$Ni$_{21}$. The line represents the fit to the measured data, 
the vertical bars denote the Bragg angles and
the bottom line shows the difference between the observed and the fitted pattern.}
\label{XRD_Hf8Ni21}
\end{figure*}

\begin{figure*}
\begin{center}
\includegraphics[scale=.65]{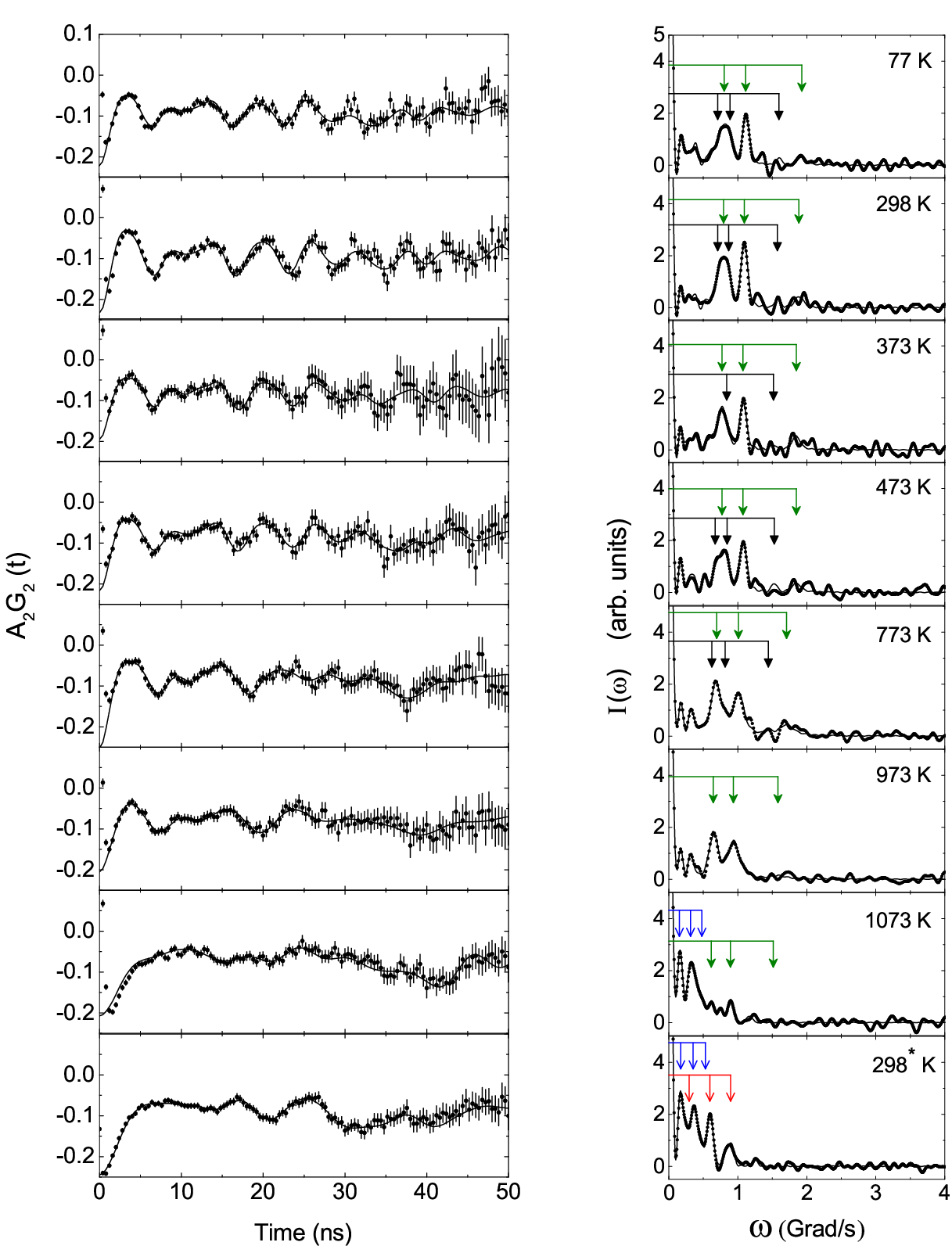}
\end{center}
\caption{\small{
TDPAC spectra in Hf$_8$Ni$_{21}$ at different temperatures. Left panel shows the time spectra and the right panel shows the
corresponding Fourier transforms. The PAC spectrum designated by 298$^*$ K is taken after the measurement at 1073 K. 
The two sets of arrows in each Fourier spectrum (up to 773 K) correspond to two
different $^{181}$Ta sites in Hf$_8$Ni$_{21}$. Two sets of arrows in the Fourier spectrum at 298$^*$ K correspond to HfNi$_3$ and Hf.}}
\label{Hf8Ni21TDPAC}
\end{figure*}

\begin{figure*}
\begin{center}
\includegraphics[scale=.25]{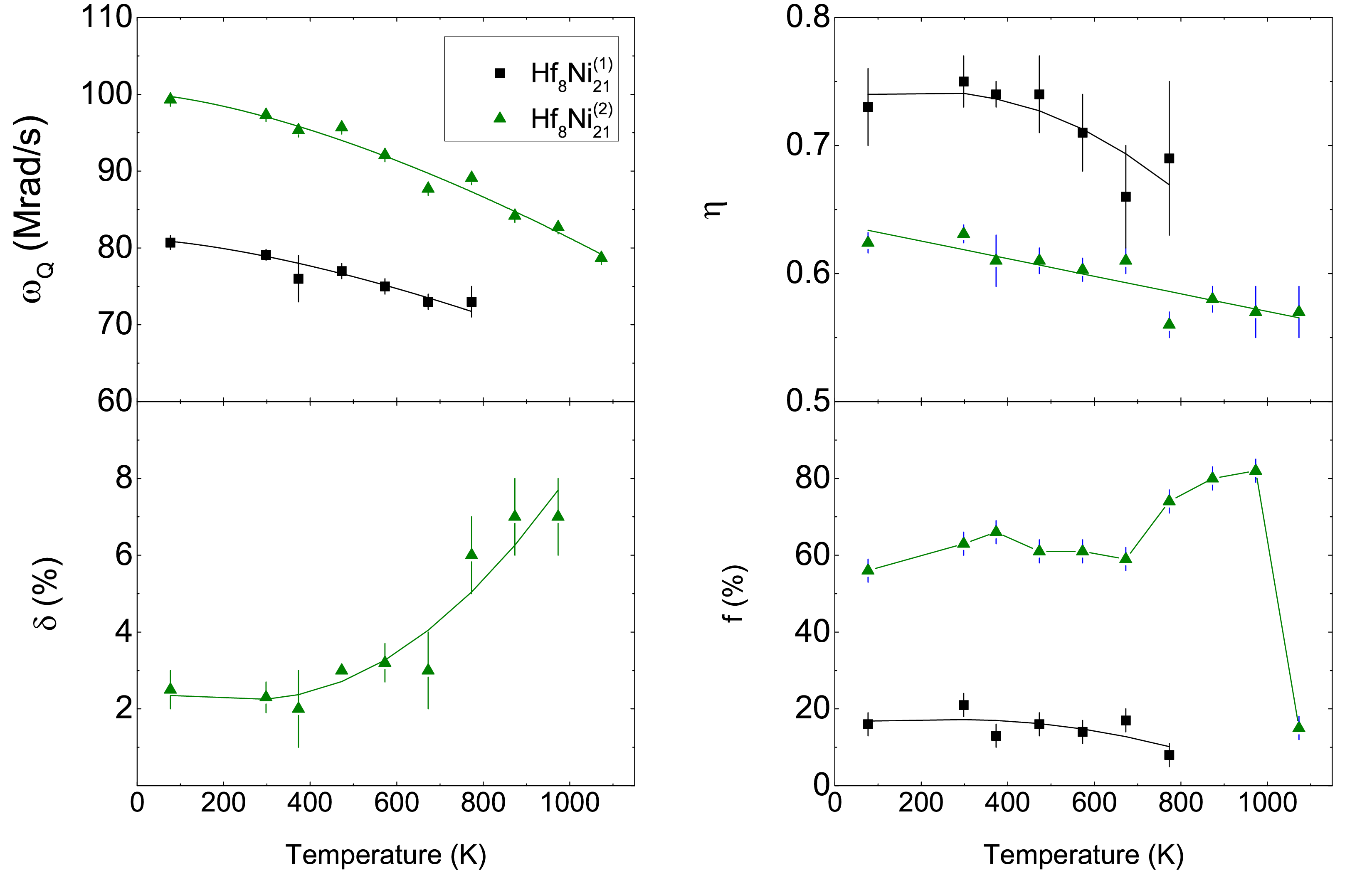}
\end{center}
\caption{Variations of quadrupole frequency ($\omega_Q$), asymmetry parameter ($\eta$) and site fraction $f$(\%) with
temperature for the two non-equivalent $^{181}$Ta sites in Hf$_8$Ni$_{21}$. 
Variations of $\delta$ is shown for the site Hf$_8$Ni$_{21}^{(2)}$.}
\label{Hf8Ni21parameter}
\end{figure*}

\begin{table}[t!]
\begin{center} 
\captionof{table}{\small{ Results of PAC measurements in Hf$_8$Ni$_{21}$}}
\scalebox{0.65}{
\begin{tabular}{ l l l l l l l l } 
 \hline  \\ [-0.9ex]
Temperature (K)  &Component    & $\omega_Q$ (Mrad/s)     & $\eta$     & $\delta$($\%$)   & $f$($\%$)      & Assignment     \\ [1.5ex]
 \hline  \\ 
 
77               &1      & 99.3(3)                 & 0.62(1)          & 2.5(5)               & 56(3)    & Hf$_8$Ni$_{21}^{(2)}$              \\ 
                 &2      & 80.7(9)                 & 0.73(3)          & 0                & 16(3)  &  Hf$_8$Ni$_{21}^{(1)}$       \\    
                 &3      & 32.3(6)                 & 0          & 0                & 19(3)  &    HfNi$_3$\\    
                 &4      & 52(2)                 & 0          & 0                & 8(3)     &  Hf     \\     \\ 

298              &1        & 97.3(3)                 & 0.63(1)          & 2.3(4)            & 63(3)      & Hf$_8$Ni$_{21}^{(2)}$       \\   
                 &2      & 79.1(7)                 & 0.75(2)          & 0                & 21(3)         & Hf$_8$Ni$_{21}^{(1)}$   \\   
                  &3      & 31(1)                 & 0               & 0                & 10(3)         & HfNi$_3$   \\   
                 &4      & 50(2)                 & 0                 & 0                & 5(3)         &Hf   \\   \\                 

373               &1     & 95.3(6)                 & 0.61(2)          & 2(1)                & 66(3)   &Hf$_8$Ni$_{21}^{(2)}$          \\   
                 &2      & 76(3)                 & 0.74(1)          & 0                & 13(3) &  Hf$_8$Ni$_{21}^{(1)}$        \\   
                 &3      & 31(2)                 & 0          & 0                & 15(3)  &  HfNi$_3$   \\
                 & 4     & 50(4)                 & 0          & 0                & 7(3)  &  Hf   \\ \\
                       
473              &1      & 95.5(5)                 & 0.62(1)          & 3(fixed)            & 61(3)   &  Hf$_8$Ni$_{21}^{(2)}$            \\ 
                 &2      & 77(1)                 & 0.74(3)          & 0                & 16(3) &  Hf$_8$Ni$_{21}^{(1)}$       \\    
                 &3      & 28.6(7)                 & 0          & 0                & 19(3) &   HfNi$_3$   \\    
                 &4      & 45(4)                 & 0          & 0                & 4(3)    &  Hf     \\     \\ 
                      
573              &1      & 92.1(3)                 & 0.60(1)          & 3.2(5)                & 61(3)    & Hf$_8$Ni$_{21}^{(2)}$            \\   
                  &2      & 75(1)                 & 0.71(3)          & 0                & 14(3) &  Hf$_8$Ni$_{21}^{(1)}$        \\    
                 &3      & 27.9(5)                 & 0          & 0                & 19(3)  &  HfNi$_3$        \\    
                 &4      & 46(2)                 & 0          & 0                & 5(3)          & Hf \\    \\ 
                       
673              &1      & 87.7(5)              & 0.61(1)          & 3(1)              & 59(3)   & Hf$_8$Ni$_{21}^{(2)}$              \\  
                 &2      & 73(1)                 & 0.66(4)          & 0                & 17(3)  &  Hf$_8$Ni$_{21}^{(1)}$        \\    
                 &3      & 27.5(9)                 & 0          & 0                & 16(3)  & HfNi$_3$     \\                     
                 &4      & 43(2)                 & 0          & 0                & 8(3)    &  Hf     \\      \\   

773              &1      & 89.1(6)                & 0.56(1)          & 6(1)                & 74(3)  &  Hf$_8$Ni$_{21}^{(2)}$           \\  
                 &2      & 73(2)                 & 0.69(6)          & 0                & 8(3) &  Hf$_8$Ni$_{21}^{(1)}$     \\    
                 &3      & 27.7(5)                 & 0          & 0                & 18(3)  & HfNi$_3$   \\    \\    
                 
873              &1      & 84.2(5)                & 0.58(1)          & 7(1)                & 80(3)    &  Hf$_8$Ni$_{21}^{(2)}$           \\  
                 &2      & 27.0(5)                 & 0          & 0                & 20(3)  & HfNi$_3$    \\    \\ 
                 
973              &1      & 82.7(7)                & 0.57(2)          & 7(1)                & 82(3)  &  Hf$_8$Ni$_{21}^{(2)}$           \\  
                 &2      & 26.8(7)                 & 0          & 0                & 18(3)  & HfNi$_3$    \\     \\
     
1073              &1      & 26.5(4)                & 0          & 5(1)                & 70(3)   &  HfNi$_3$           \\  
                 &2      & 78.7(6)                 & 0.57(2)          & 0                & 15(3)  & Hf$_8$Ni$_{21}^{(2)}$  \\    
                 &3      & 63.3(9)                 & 0.23(8)          & 0                & 15(3)  & Hf$_2$Ni$_7$   \\    \\ 
                            
298$^*$           &1   & 30.2(4)                 & 0          &9(2)                & 62(3)       & HfNi$_3$  \\ 
                  &2   & 50.8(4)                & 0          & 0              & 25(3)        & Hf \\ 
                  &3   & 70.0(8)                & 0          & 0              & 12(3)        & Hf$_2$Ni$_7$     \\                                              
\hline
\end{tabular}}
\begin{flushleft}
 \small{$^*$ after measurement at 1073 K}\\
\end{flushleft}
\label{tab:Hf8Ni21table}
\end{center}
\end{table}

\subsection{Zr$_8$Ni$_{21}$} 
The XRD powder pattern of Zr$_8$Ni$_{21}$ sample is shown in figure $\ref{XRD_Zr8Ni21}$. The X-ray analysis has been
carried out using the known crystallographic data of Zr$_8$Ni$_{21}$ \citep{JM}. 
The XRD spectrum shows no other peaks except for 
Zr$_8$Ni$_{21}$ and this sample is, therefore, found to be produced in an almost pure single component phase. If any small
contaminating phases like Zr$_7$Ni$_{10}$ or Zr$_2$Ni$_7$ are produced, it is not observed from the XRD powder pattern.

The TDPAC spectra of $^{181}$Ta in the as prepared sample of Zr$_8$Ni$_{21}$ are shown in figure \ref{Zr8Ni21TDPAC}. From PAC
measurements, three frequency components at room temperature have been obtained (Table \ref{tab:Zr8Ni21table}). 
The components 1 and 3 have been attributed to Zr$_8$Ni$_{21}$ by 
comparing with our PAC results in Hf$_8$Ni$_{21}$ and results from ab-initio calculations by density functional theory (discussed later). The component 2 has been
attributed to Zr$_7$Ni$_{10}$. This follows from the previous X-ray EDS and SEM/TEM results reported by Shen et al. \citep{Shen}. 
The characteristic frequency and $\eta$ for 
this component are distinctly different than those found in Zr$_2$Ni$_7$ \citep{CCDeyPhysica}. Moreover,
assignment of this component can be supported 
from our PAC measurements in Zr$_7$Ni$_{10}$ \citep{skdeyJAC} where, a component similar to this was found. It is found that
   three frequency components are
   required to fit the spectra in the temperature range 77-373 K with no appreciable
   change in parameters (Table \ref{tab:Zr8Ni21table}). Variations of $\omega_Q$, $\eta$ and site fraction ($f$) with temperature 
   for the two components of Zr$_8$Ni$_{21}$ are shown in figure \ref{Zr8Ni21parameter}.
   
  At 473 K, however, the PAC spectrum gives two components. The component Zr$_8$Ni$_{21}^{(2)}$ does not exist at this temperature.  In the temperature
  range 473-973 K, no appreciable changes in results are observed (Table \ref{tab:Zr8Ni21table}). But at 1073 K, a drastic change in PAC spectrum is found. At
  this temperature, 
  the predominant component ($\sim$68\%) produces a sharp decrease in quadrupole frequency and asymmetry parameter shows a value equal to zero. This possibly 
  indicates a change in local environment of the probe. The component due to Zr$_7$Ni$_{10}$, on the other hand, remains almost unchanged.
  
  To understand the change in PAC spectrum at 1073 K, we have repeated the PAC measurement at room temperature. 
  In the re-measured
  spectrum at 298 K, the predominant component ($\sim$70$\%$) produces values of $\omega_Q\sim$57 Mrad/s, $\eta\sim$0. At the remeasured room temperature, it is found 
  that the major component of Zr$_8$Ni$_{21}$ found initially at room temperature reappears with a much smaller fraction (Table \ref{tab:Zr8Ni21table}). One additional 
  new component is observed with a very small fraction which can be assigned to Zr$_2$Ni$_7$ by comparing with the previous result
  in Zr$_2$Ni$_7$ \citep{CCDeyPhysica}. From SEM/X-ray EDS measurement \citep{Shen} also, 
Zr$_2$Ni$_7$ was found in a sample of Zr$_8$Ni$_{21}$ annealed at 1233 K.

We have performed XRD measurement in a sample of Zr$_8$Ni$_{21}$ annealed at 1073 K for two days. The XRD spectrum
(Fig. \ref{XRD_Zr8Ni21}) shows peaks mainly due to Zr$_8$Ni$_{21}$. This indicates that no major structural or compositional phase transformation occurs at 1073 K.

Possible explanation for the predominant component observed at 1073 K and subsequently at room temperature is the following. Probably, the probe $^{181}$Hf 
was not settled well at the position of Zr$_8$Ni$_{21}$ and at 1073 K, these probe atoms got enough energy  to go out from the position. The major component is, 
therefore, observed due to the Hf probe itself. 
  
The electric field gradients in metal and intermetallic compound are found to vary with temperature following $T$ or $T^{3/2}$ relationship \citep{Christiansen}. In Zr$_8$Ni$_{21}$,
it is found 
that quadrupole frequencies vary with $T^{3/2}$ for both components. 
For the
predominant site Zr$_8$Ni$_{21}^{(1)}$ (present up to 973 K), the results are fitted by
  \begin{equation}
   \omega_Q(T)=\omega_Q(0)[1-\beta T^{3/2}]. 
   \label{eqn:T32}
  \end{equation}
 A least squares fitting gives results $\omega_Q$(0)=78.2(1) Mrad/s and $\beta$=5.9(1)$\times$10$^{-6}$ K$^{-3/2}$.  
  
  \begin{figure*}
\begin{center}
\includegraphics[scale=.7]{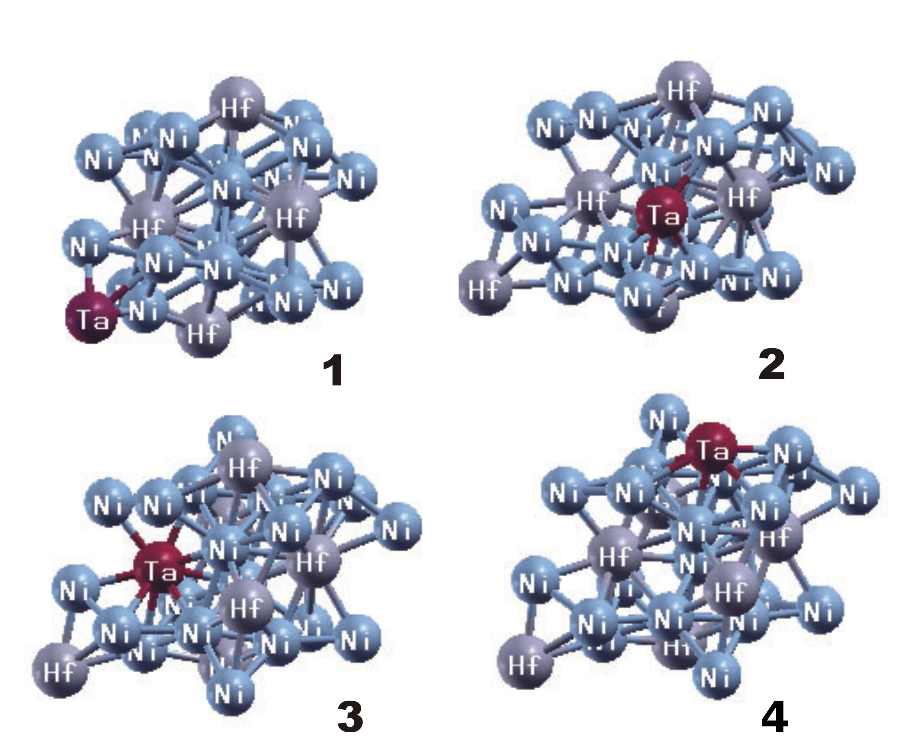}
\end{center}
\caption{Models of four types of cells used in this study}
\label{Atomic_position_Hf8Ni21}
\end{figure*}
\begin{table*}[t!]
\begin{center} 
\captionof{table}{\small{  Calculated EFG values in units of 10$^{21}$ V/m$^2$ and asymmetry parameters}}
\scalebox{0.70}{
\begin{tabular}{ l l l l l l l } 
 \hline  \\ [-0.9ex]
Probe  &Lattice Site    & EFG  & $\eta$ & Measured EFG \\&&&&extrapolated to 0 K  &Measured $\eta$ at 77 K  \\ [1.5ex]
 \hline  \\ 
 
no probe (pure compound)               &Zr(1) 2i 0.0691(4) 0.9052(3) 0.3147(3)      & -3.5                & 0.82                  \\ 
          Zr$_8$Ni$_{21}$       &Zr(2) 2i 0.2460(4) 0.4364(3) 0.0344(3)      & -4.1                 & 0.85                \\    
                 &Zr(3) 2i 0.2462(4) 0.5604(3) 0.6070(3)      & -5.0                & 0.88             \\    
                 &Zr(4) 2i 0.5685(4) 0.0374(3) 0.2475(3)      & -4.7                 & 0.74                 \\     \\ 

$^{181}$Ta in Zr$_8$Ni$_{21}$              &Zr(1) 2i 0.0691(4) 0.9052(3) 0.3147(3)    &8.8    & 0.91  & 8.7(2)  & 0.80(1)                 \\   
                 &Zr(2) 2i 0.2460(4) 0.4364(3) 0.0344(3)      & -11.4                 & 0.55           \\   
                  &Zr(3) 2i 0.2462(4) 0.5604(3) 0.6070(3)      & -12.5                 & 0.78    &11.4(1)  &   0.73(3)        \\   
                 &Zr(4) 2i 0.5685(4) 0.0374(3) 0.2475(3)      & -12.3                 & 0.56               \\   \\                 

no probe (pure compound)  &Hf(1) 2i 0.2487(3) 0.0620(3) 0.1072(3)     & -11.5                & 0.75                 \\ 
      Hf$_8$Ni$_{21}$   &Hf(2) 2i 0.0688(3) 0.4040(2) 0.8150(2)      &-8.1                 & 0.91                \\    
                 &Hf(3) 2i 0.4312(3) 0.4633(3) 0.2543(3)      & -10.8                & 0.69             \\    
                 &Hf(4) 2i 0.2452(3) 0.9375(3) 0.5340(2)     & -10.2                 & 0.64                 \\     \\ 

$^{181}$Ta in Hf$_8$Ni$_{21}$              &Hf(1) 2i 0.2487(3) 0.0620(3) 0.1072(3)   &-13.3     & 0.71                    \\   
                 &Hf(2) 2i 0.0688(3) 0.4040(2) 0.8150(2)      & 9.6                 & 0.88  &9.1(1) &  0.73(3)        \\   
                  &Hf(3) 2i 0.4312(3) 0.4633(3) 0.2543(3)      & -12.5                 & 0.60                 \\   
                 &Hf(4) 2i 0.2452(3) 0.9375(3) 0.5340(2)      & -11.8                & 0.51 &11.2(2) & 0.62(1)             \\   \\  
                                              
\hline
\end{tabular}}
\label{tab:DFT_Zr8Ni21_Hf8Ni21}
\end{center}
\end{table*}
    \subsection{Hf$_8$Ni$_{21}$}
    The powder XRD pattern of Hf$_8$Ni$_{21}$ is shown in figure \ref{XRD_Hf8Ni21}. The spectrum is fitted using the known crystallographic 
    data of Hf$_8$Ni$_{21}$ \citep{Bsenko}. The X-ray analysis shows that there are no other peaks except for Hf$_8$Ni$_{21}$ and this sample is also found to be
    produced in an almost pure single 
    component phase. No contaminating phase is obtained from the XRD powder pattern.
    
    The TDPAC spectra of $^{181}$Ta in Hf$_8$Ni$_{21}$ are shown in figure \ref{Hf8Ni21TDPAC}. The spectrum
   at room temperature produces four interaction frequencies.
   The first two components at room temperature with site fractions 63\% and 21\%
   (Table \ref{tab:Hf8Ni21table}) are found to be quite similar to the components found in Zr$_8$Ni$_{21}$. 
   The third weak component
   can possibly be attributed to HfNi$_3$ produced along with
   Hf$_8$Ni$_{21}$. L. Bsenko \citep{LARS BSENKO} reported
   the decomposition of Hf$_8$Ni$_{21}$ to HfNi$_3$ eutectoidally at 1175$\pm$10$^\circ$C. This component has been found in the
   whole temperature range. The assignment of HfNi$_3$ in Hf$_8$Ni$_{21}$ can be supported also from our PAC measurements in HfNi$_3$ \citep{skdeyCCDey}
   where a similar component to this was found. A very weak component ($\sim$5\%) found at room temperature can be 
   attributed to Hf probe which is not settled in the compound.
 From our temperature dependent PAC measurements, it is found that all
 four components exist in the temperature range 77-673 K. The component 4
 is not observed at 773 K and the minor site of Hf$_8$Ni$_{21}$ (Hf$_8$Ni$_{21}^{(1)}$) disappears
 at 873 K. A drastic change in PAC spectrum is observed at 1073 K
 where the tentatively assigned HfNi$_3$ component suddenly increases at the expense of Hf$_8$Ni$_{21}$ (Table \ref{tab:Hf8Ni21table}). The component
 due to Hf$_8$Ni$_{21}$ reduces to only 15\%.
 At this temperature, a new
 frequency component (component 3) is observed which probably can be attributed to Hf$_2$Ni$_7$ by comparing its values with the results reported in the analogous compound Zr$_2$Ni$_7$ \citep{CCDeyPhysica}. 
 
 After measurement at 1073 K, a re-measurement
 at 298 K is carried out. In the re-measurement, HfNi$_3$ is found to be predominant ($\sim$62\%) 
 which appeared as a minor fraction ($\sim$10\%) initially at room temperature. Here, no component due to Hf$_8$Ni$_{21}$ 
 is observed. A small component fraction of Hf$_8$Ni$_{21}$ found at 1073 K and absence of this fraction at remeasured room temperature indicates that
 Hf$_8$Ni$_{21}$ is not a stable phase approximately above 1000 K. It is found
 also that, the component due to Hf probe atom re-appears at this temperature with a higher component fraction ($\sim$25\%). The quadrupole frequency and asymmetry
 parameter for this component ($\omega_Q$=50.8 Mrad/s, $\eta$=0) are very much similar to the values in a Hf metal \citep{SKDEYJPCS}.
  
  Variations of $\omega_Q$, $\eta$, $\delta$ and site fraction ($f$) for different
  components observed in Hf$_8$Ni$_{21}$ in the temperature range 
  77-1073 K are shown in figure \ref{Hf8Ni21parameter}. In Hf$_8$Ni$_{21}$ also,
  the quadrupole frequencies for the two sites show $T^{3/2}$
  temperature dependent behaviors. A least squares fitting (eqn. \ref{eqn:T32}) for the predominant site ( present up to 1073 K) 
  gives values of $\omega_Q$(0)=100.1(5) Mrad/s, $\beta$=6.0(3)$\times$10$^{-6}$ K$^{-3/2}$. For the minor site (present up to 773 K), the fitted results are found 
  to be $\omega_Q$(0)=81.2(4) Mrad/s, $\beta$=5.4(3)$\times$10$^{-6}$ K$^{-3/2}$. Similar values of $\beta$ for both components indicate similar temperature 
  dependent behaviors for the two sites. Also, the values of $\beta$ are quite similar to the value of $\beta$ in Zr$_8$Ni$_{21}$ for the site Zr$_8$Ni$_{21}^{(1)}$.

\section{DFT calculation}
The first-principles density functional theory
(DFT) calculations were performed to compare with the
experimental results and to dispel the doubts existing in
the interpretation of the experimental data. All the
calculations were done with WIEN2k simulation package \citep{Blaha}, based on the full potential (linearized) augmented
plane waves method (FP (L)APW). Electronic exchange-correlation energy was treated with generalized gradient
approximation (GGA) parametrized by Perdew-Burke-Ernzerhof (PBE) \citep{Perdew}. In our calculations the muffin-tin
radii for Zr, Ni and Ta (Hf) were 2.3, 2.2 and 2.4 a. u., 
respectively. The cut-off parameter $R_{mt}K_{max}$ for limiting
the number of plane waves was set to 7.0, where $R_{mt}$ is
the smallest value of all atomic sphere radii and $K_{max}$ is
the largest reciprocal lattice vector used in the plane
wave expansion.

The Brillouin zone integrations within the self-consistency cycles were performed via a tetrahedron
method \citep{Blochl}, using 18 $k$ points in the irreducible wedge
of the Brillouin zone (4$\times$3$\times$3 mash). The atomic
positions were relaxed according to Hellmann-Feynman forces calculated at the end of each self-consistent cycle,
with the force minimization criterion 2 mRy/a.u.. In our calculations the self-consistency was
achieved by demanding the convergence of the
integrated charge difference between last two
iterations to be smaller than 10$^{-5}$. Both Zr$_8$Ni$_{21}$ and Hf$_8$Ni$_{21}$ crystallize in the triclinic P1 type structure,
which possesses 15 non-equivalent crystallographic
positions \citep{JM,Bsenko}, 4 for Zr (Hf) atoms and 11 for Ni
atoms. 
All Zr (Hf) non-equivalent positions have the same point group symmetry
2$i$ and 3 Zr and 12 Ni atoms as nearest neighbors, except Zr(3),
which has 2 Hf and 13 Ni. Each of the four non-equivalent Zr (Hf)
atoms in the unit cell, stated in references \citep{JM,Bsenko} was replaced by Ta subsequently (figure \ref{Atomic_position_Hf8Ni21}, \citep{Kokalj}) preserving the point
group symmetry around original atom and then electric
field gradients at thus created Ta positions were
calculated using the method developed in reference \citep{BlahaPRL}.

The usual convention is to designate the largest
component of the EFG tensor as $V_{zz}$. The asymmetry
parameter $\eta$ is then given by $\eta=(V_{xx}-V_{yy})/V_{zz}$, where $V_{zz}\ge V_{yy}\ge V_{xx}$. The calculated EFGs in the
pure compounds as well as at Ta probe positions in the
investigated compounds are given in Table \ref{tab:DFT_Zr8Ni21_Hf8Ni21}.

It can be observed that there is not much difference in
the EFG values for four non-equivalent Zr positions in
the pure Zr$_8$Ni$_{21}$ compound. EFG is the smallest at Zr1 and
the largest at Zr3. This trend is preserved also for the
electric field gradients calculated at the corresponding
Ta positions, but the EFGs are now about 2.5 times
larger. In the pure Hf$_8$Ni$_{21}$ compound, EFG values are
about doubled, as compared to the corresponding
ones for Zr$_8$Ni$_{21}$, but the $\eta$ values are similar. Here,
also, introduction of Ta atom at one of the non-equivalent Hf sites, leads to increased EFG values.

\section{Discussion}
In the temperature range 77-373 K, Zr$_8$Ni$_{21}$ PAC spectra
consist of three frequency components. A uniform conversion
from the measured quadrupole frequencies to the EFGs is
achieved using the value of 2.36$\times$10$^{-24}$ cm$^2$ \citep{Butz} for the
quadrupole moment of $^{181}$Hf. By comparing the measured
results for the EFGs and asymmetry parameters with the
calculated ones, components 1 and 3 (Table \ref{tab:Zr8Ni21table}) are attributed
to the two non-equivalent Zr sites in Zr$_8$Ni$_{21}$. The measured values of
EFGs ( 8.7$\times$10$^{21}$ V/m$^2$ and 11.3$\times$10$^{21}$ V/m$^2$) and $\eta$ (0.80 and
0.73) at 77 K are in
excellent agreement with the calculated values for the two Zr
sites in Zr$_8$Ni$_{21}$. However, as Ta doped Zr$_8$Ni$_{21}$ has four non-equivalent crystallographic positions
with similar EFG and asymmetry parameter (Table \ref{tab:DFT_Zr8Ni21_Hf8Ni21}), in order to explain
preferential site occupation, we performed ab initio total energy calculations
for Ta doped Zr$_8$Ni$_{21}$ and found that the configuration obtained when Ta replaces
Zr(3) position has the lowest formation energy, about 0.013 eV lower than the
structure when Ta is at Zr(1) postion. The formation energies of the remaining two configurations are about 0.1 eV higher.

At 1073 K, there is a drastic change of PAC spectrum in Zr$_8$Ni$_{21}$. At this temperature, EFG for the predominant component produces
a zero value of $\eta$. A similar change in $^{181}$Ta PAC spectra with increasing temperature
above 650 K was observed in TiPd$_2$ compound \citep{BWodniecki} and was explained with the shift
of Ta atom from Ti to Pd lattice site \citep{BWodniecki,Cavor}, but in our case, DFT calculations
excluded that possibility, as all of the non-equivalent Ni sites in Zr$_8$Ni$_{21}$ has $\eta$ which differs from zero. 
We find a resemblance of EFG and $\eta$ for the component 1 in Zr$_8$Ni$_{21}$
at re-measured room temperature and component 4 in Hf$_8$Ni$_{21}$ 
with the calculated values for Ta in pure Hf metal (6.7$\times$10$^{21}$ V/m$^2$ and $\eta$ = 0). 

The PAC spectrum for Hf$_8$Ni$_{21}$ at room temperature consists of four components.
The first two components with the EFG values 11.1$\times$10$^{21}$ V/m$^2$ and 9.0$\times$10$^{21}$ V/m$^2$
and the corresponding asymmetry parameters 0.62 and 0.73 (at 77 K) obviously correspond
to the two different Hf positions in Hf$_8$Ni$_{21}$ (Table \ref{tab:DFT_Zr8Ni21_Hf8Ni21}). The measured results show
that quadrupole frequencies for the two corresponding sites in Zr$_8$Ni$_{21}$ and Hf$_8$Ni$_{21}$
vary in similar manner with temperature.  

\section{Conclusion} 
 We have presented the time differential perturbed angular correlations
 measurements and DFT calculations to determine the electric field gradients in Zr$_8$Ni$_{21}$ and Hf$_8$Ni$_{21}$ intermetallic
 compounds. Our results indicate that during the preparation of Zr$_8$Ni$_{21}$ by arc
 melting, other phases like Zr$_7$Ni$_{10}$ can also be formed. The same goes for Hf$_8$Ni$_{21}$,
 in which HfNi$_3$ compound was detected.  In both Zr$_8$Ni$_{21}$  and Hf$_8$Ni$_{21}$, EFGs for two
 non-equivalent sites of Zr/Hf, vary following $T^{3/2}$ relationship with temperature. Temperature dependent PAC measurement show that Hf$_8$Ni$_{21}$ is 
 probably not a stable phase above 1000 K.
  
  \newpage
 {\hspace{ -0.4 cm}}{\bf Acknowledgement}
 \vspace{0.5cm}
 
One of the authors (C.C. Dey) gratefully acknowledges the
help of Prof. Dr. T. Butz, University of Leipzig, Germany in data
analysis. The authors acknowledge with thanks A. Karmahapatra and S. Pakhira of Saha Institute of Nuclear Physics, Kolkata for their helps in XRD
measurements and data analysis. The present work is supported by the Department of Atomic Energy, Government of India through the Grant
no. 12-R\&D-SIN-5.02-0102. Finally, J. Belo$\check{\text{s}}$evi\'{c}-$\check{\text{C}}$avor acknowledges support by The Ministry of Education, 
Science and Technological Department of the Republic of Serbia through the project no. 171001.

\bibliographystyle{elsarticle-num-names}

\begin{thebibliography}{9}
 
 \bibitem{JMJoubert} J.M. Joubert, M. Latroche, A. Percheron-Gu\`{e}gan, J. Alloys Compd. 231 (1995) 494. 
 
 \bibitem{Ruiz} F.C. Ruiz, E.B. Castro, S.G. Real, H.A. Peretti, A. Visintin, W.E. Triaca, Int. J. Hydrogen Energy, 33 (2008) 3576.
 
 \bibitem{FCRuiz}F. C. Ruiz, E. B. Castro, H. A. Peretti, A. Visitin, Int. J. Hydrogen Energy 35 (2010) 9879.
 
\bibitem{Regmi} J. Nei, K. Young, R. Regmi, G. Lawes, S.O. Salley, K.Y.S. Ng, Int. J. Hydrogen Energy 37 (2012) 16042. 

\bibitem{S.O.Salley} J. Nei, K. Young, S. O. Salley, K. Y. S. Ng, J. Alloys Compd. 516 (2012) 144.

\bibitem{Young} Kwo-hsiung Young, Jean Nei, Materials 6 (2013) 4574.

\bibitem{Drulis}H. Drulis, W. Iwasieczko, V. Zaremba, J. Mag. Mag. Materails 256 (2003) 139.

\bibitem{Amamou}A. Amamou, Solid State commun. 37 (1981) 7.

\bibitem{Dey}C. C. Dey, J. Mag. Mag. Materials 342 (2013) 87.

\bibitem{SilvaJMMM} P.R.J. Silva, H. Saitovitch, J.T. Cavalcante, M. Forker, J. Mag. Mag. Materials 322 (2010) 1841.

\bibitem{CCDeyPhysica} C. C. Dey and S. K. Srivastava, Physica B 427 (2013) 126.

 \bibitem{CCDeyJPCS} C. C. Dey, Rakesh Das and S. K. Srivastava, J. Phys. Chem. Solids 82 (2015) 10.

\bibitem{Shen}Haoting Shen, Leonid A. Bendersky, Kwo Young, Jean Nei, Materials 8 (2015) 4618.

\bibitem{JM}J. M. Joubert, R. cern\`{y}, K. Yvon, Z. Kristallogr. New Cryst. Struct. 213 (1998) 227.

\bibitem{Bsenko} L. Bsenko, Acta. Cryst. B 34 (1978) 3204.

\bibitem{Schatz} G. Schatz, A. Weidinger, Nuclear condensed matter physics, Nuclear Methods and Application, John Wiley and Sons,
Chichester, New York, Brisbane, Toronto, Singapore, 1996, p. 63 (chapter 5).

\bibitem{Catchen}G. L. Catchen, Mater. Res. Soc. Bull. XX (7), 37 (1995).

\bibitem{Zacate}M. Zacate and H. Jaeger, Defect Diffus. Forum 311 (2011) 3.

\bibitem{Errico}L. A. Errico, H. M. Petrilli, L. A. Terrazos, A. Kuli\'{n}ska,
P. Wodniecki, K. P. Lieb, M. Uhrmacher, J. Belosevic-Cavor, V. Koteski,
J. Phys. Condens. Matter 22 (2010) 215501.

\bibitem{Wodniecki}P. Wodniecki, A. Kuli\'{n}ska, B. Wodniecka, S. Cottenier, H. M. Petrilli, M. Uhrmacher, K. P. Lieb, Europhysics Letters 77 (2007) 43001.

  \bibitem{Firestone} R. B. Firestone, V. S. Shirley (Eds.), Table of Isotopes, 8th ed., John Wiley and Sons, New York, 1996.

   \bibitem{pramana} C. C. Dey, Pramana 70 (2008) 835.
   
   \bibitem{skdeyJAC} S. K. Dey, C. C. Dey, S. Saha, J. Belo$\check{\text{s}}$evi\'{c}-$\check{\text{C}}$avor (to be published).
      
   \bibitem{Christiansen} W. Witthuhn and E. Engel, in Hyperfine Interactions of Radioactive Nuclei, Ed. by J. Christiansen (Springer-Verlag
   Berlin Heidelberg New York Tokyo, 1983) P. 205.
      
   \bibitem{LARS BSENKO} Lars Bsenko, J. Less Common Met. 63 (1979) 171.
   
   \bibitem{skdeyCCDey} S. K. Dey, C. C. Dey, S. Saha, J. Belo$\check{\text{s}}$evi\'{c}-$\check{\text{C}}$avor (to be published).
   
         \bibitem{SKDEYJPCS} S. K. Dey, C. C. Dey, S. Saha, J. Phys. Chem. Solids 95 (2016) 98.
     
\bibitem{Blaha}P. Blaha, K. Schwarz, G. K. H. Madsen, D. Kvasnicka, J. Luitz, WIEN2k an Augmented Plane Wave Plus Local Orbitals Program for 
       Calculating Crystal Properties, Vienna University of Technology,
Vienna, Austria, 2001.

\bibitem{Perdew} J. P. Perdew, K. Burke, M. Ernzerhof, Phys. Rev. Lett. 77 (1996) 3865.

\bibitem{Blochl} P. E. Blochl, O. Jepsen, O. K. Andersen, Phys. Rev. B 49 (1994) 16223.

\bibitem{Kokalj} A. Kokalj, J. Mol. Graphics Modelling 17 (1999) 176.

\bibitem{BlahaPRL}P. Blaha, K. Schwarz, Phys. Rev. Lett. 54 (1985) 1192.

\bibitem{Butz}T. Butz, A. Lerf, Phys. Lett. A 97 (1983) 217.

\bibitem{BWodniecki} P. Wodniecki, B. Wodniecka, A. Kuli\'{n}ska, M. Uhrmacher, K. P. Lieb, Journal of Alloys and Compounds 385 (2004) 53.

\bibitem{Cavor} J. Belo$\check{\text{s}}$evi\'{c}-$\check{\text{C}}$avor, V. Koteski, J. Radakovi\'{c}, Solid State Communications, 152 (2012) 1072. 

\end{thebibliography}

\end{document}